\documentclass[10pt,showpacs,letterpaper,aps,pra,twocolumn,notitlepage,superscriptaddress]{revtex4-1}

\usepackage{amssymb,amsmath,amsfonts,amsthm,bbm}
\usepackage{multirow}
\usepackage{booktabs}
\usepackage{graphicx}

\DeclareMathOperator{\re}{\mathbb{R}e}
\DeclareMathOperator{\im}{\mathbb{I}m}
\def\realpart#1{{\re \! \left[ #1 \right]}}
\def\imagpart#1{{\im \! \left[ #1 \right]}}

\def\tA{\widetilde{A}}
\def\tw{\widetilde{w}}
\def\tsigma{\widetilde{\sigma}}
\def\tR{\widetilde{R}}

\def\Gaussian#1#2{{N^{#1}\left(#2\right)}}
\def\GaussianNoParm#1{{N^{#1}}}

\def\IdapproxI{\overline{I}_1}
\def\IdapproxII{\overline{I}_2}
\def\Gdapprox{\overline{G}_{d}}

\def\gdapprox{\overline{g}_{d}}

\def\GradConst{{C}}
\def\GradMain#1{{\Psi_{#1}}}
\def\GradMainNorm#1{{\widetilde{\Psi}_{#1}}}
\def\GradGauss#1{{Z_{#1}}}

\def\GradGdapproxExp{{D}}

\def\GradNorm{{P}}
\def\GradNormApprox{{\overline{P}}}

\def\MaxGerror{\max \{f\}}
\def\GerrorGeneral{f}
\def\Gerror{f}
\def\gerror{\widetilde{f}}
\def\GerrorSquaredSum{F}

\def\dGerrorSquaredSumda{\frac{\partial \GerrorSquaredSum}{\partial \avec}}

\def\SampleCount{M}
\def\sampleyI#1{y_1^#1}
\def\sampleyII#1{y_2^#1}

\def\StepSizBreakpoint{\mu_\avec}
\def\StepSizDistance{\mu_d}

\def\avec{{\mathbf{a}}}

\def\aveciter#1{{\avec^{(#1)}}}
\def\diter#1{{d^{(#1)}}}

\def\ThresholdGradBreakpoint{\tau_\avec}
\def\ThresholdGradDistance{\tau_d}

\def\ThresholdError{\epsilon}
\def\ThresholdMeanError{\gamma}
\def\ThresholdErrorVotes{p}
\def\ErrorVoteSet{\kappa}

\def\TwoObjectGdapproxInverted#1{\overline{G}^{#1}_{d}}
\def\TwoObjectSum{\breve{G}_{d}}
\def\TwoObjectParam#1#2{{#1^#2}}

\newtheorem*{thmnonum}{Theorem}

\begin{document}

\title{On the inverse problem of source reconstruction from coherence measurements}

\author{Andre Beckus}
\affiliation{Department of Electrical and Computer Engineering, University of Central Florida, Orlando, FL 32816, USA}
\author{Alexandru Tamasan}
\affiliation{Department of Mathematics, University of Central Florida, Orlando, FL 32816, USA}
\author{Aristide Dogariu}
\affiliation{CREOL, The College of Optics \& Photonics, University of Central Florida, Orlando, FL 32816, USA}
\author{Ayman F. Abouraddy}
\affiliation{CREOL, The College of Optics \& Photonics, University of Central Florida, Orlando, FL 32816, USA}
\author{George K. Atia}
\affiliation{Department of Electrical and Computer Engineering, University of Central Florida, Orlando, FL 32816, USA}

\begin{abstract}
We consider an inverse source problem for partially coherent light propagating in the Fresnel regime. The data is the coherence of the field measured away from the source. The reconstruction is  based on a minimum residue formulation, which uses the authors' recent closed-form approximation formula for the coherence of the propagated field. 
The developed algorithms require a small data sample for convergence and yield stable inversion by exploiting information in the coherence as opposed to intensity-only measurements.
Examples with both simulated and experimental data demonstrate the ability of the proposed approach to simultaneously recover complex sources in different planes transverse to the direction of propagation.
\end{abstract}

\maketitle

\section{Introduction}
The reconstruction of a source of light from measured field data is one of the central problems in optics, with applications ranging from microscopy to astronomy. Traditional methods are mostly based on intensity measurements (see e.g., \cite{George97OC} and references therein).
At the same time, it is well known that the spatial coherence function is an excellent encoder of information (such as location, spatial extent, etc.) about the source: for the simple cases involving apertures we refer to Section 5.7 of \cite{Goodman:85}, and for the quasi-homogeneous partially coherent sources to \cite{COLLETT198027}; see also \cite{Kondakci17OE,ElHalawany17} for more complex source configurations.
Here, we exploit this information to characterize a source using coherence measurements.
Our work is particularly relevant for geometries where the shadow of an object is not informative or for situations where the source of light is unresolved and the distance to the source is also of interest. Irrespective of the domain of operation across the electromagnetic spectrum, these are circumstances characterized by small Fresnel numbers where the size and location of objects cannot be simply determined from a planar distribution of intensity.
An added benefit in using coherence data is the overdeterminancy of the problem which we exploit to develop a robust inversion method.

Despite an abundance of methods for measuring the coherence function (from classical double slit approaches \cite{Thompson:57,DivittNonParallelSlits} to modern slit realization using digital micromirrors \cite{CoherenceDMD,Kondakci17OE,ElHalawany17}, to shearing interferometers \cite{Iaconis96OL,Cheng00JMO,RezvaniNaraghi:17} and microlens arrays \cite{Stoklasa14NC}), there are very few works that use this data to recover the source. 
For non-radiating sources, unique determination ideas appeared in \cite{LaHaie:86,GBUR2001301}. In the Fresnel regime, some Fourier-based inversion methods use the van Cittert-Zernike theorem to recover the intensity distribution across
incoherent sources \cite{FundOfPhotonics}, and the more complicated case of partially coherent quasi-homogeneous sources \cite{Garter:85,doi:10.1063/1.524277,LaHaie:85}.  
Further algorithms use only the modulus of the Fourier transform \cite{Kohler:73,Fienup78OL}, with various extensions (e.g., the use of apriori constraints  \cite{Fienup:87} or coherent illumination \cite{Fienup:06}) which improve the reconstruction.
However, the accuracy of these methods degrades with the increase in the coherence of the source. 
In the near-field regime, a  successful method
reconstructs complex sources by the back-propagation of the measured coherence function \cite{ElHalawany17}.
Other means of inversion are based on coherent modes \cite{0266-5611-13-1-005} or Fresnelets \cite{1187353}.

While the above-mentioned inversion methods allow for the estimation of arbitrary intensity profiles, in practice, they all suffer from large sampling complexity. Specifically, in order to invert a Fourier or Fresnel transform, a large number of measurements is necessary to attain the required sampling rate.
In the back-propagation approach (which requires the full coherence function), the source is traced back in an increasing sequence of distances away from the measurement plane; hence, the reconstruction requires identification of the correct axial distance. This information is typically unavailable or hard to obtain. Even if the distance is identified, all calculations at the intermediate locations would then be discarded, which adds an unnecessary computational expense. 

In this paper, we present an inversion method to reconstruct sources from coherence measurements, while avoiding the aforementioned pitfalls.
We exploit the additional dimension in the coherence data to devise a global inversion method that applies local minimization to a family of residuals sharing a unique minimum, a task that would be difficult from intensity-only measurements.
In our previous work \cite{Beckus:17}, we studied the propagation of the spatial coherence of fields from \emph{generalized sources} in the Fresnel regime. Such sources are modulations of the field produced by a Gauss-Schell source by a piecewise constant transmission function, thus modeling the field's interaction with objects and apertures. 
We adopt this formulation due to the analytical tractability of its forward model and its applicability to many practical scenarios of interest. Our focus here is on the inverse problem in which we seek to determine both the transmission function and the distance to the generalized source from the measurement plane from sampled coherence measurements. 
Leveraging the closed-form approximations obtained in \cite{Beckus:17} (which are explicit in the parameters of the transmission function) along with parametric modeling of the scene, we develop a gradient-descent-based approach to the inverse problem. The proposed algorithm yields accurate estimates of the parameters of the scene with low sampling complexity, i.e., only few measured samples of the coherence function suffice for the algorithm to converge to the actual parameters.   
We limit this study to the one-dimensional model, in which the field is assumed to vary only along one transverse direction. However, the techniques developed here are extendable to higher dimensions.
While we focus on intercepting objects, which obstruct part of the light source, the method applies to more complex source structures as in \eqref{eqn:generalized_source} below. In particular, the complementarity in the Babinet principle for mutual intensity \cite{Sukhov:17}, directly allows the method to apply to secondary sources or apertures. 

We start with a simple model involving one source. The goal is to estimate its position, width, and distance from coherence measurements. Inversion using both the numerically simulated, and experimental data are presented to demonstrate the algorithm's effectiveness. A second example considers two sources whether located in the same or in different transverse planes.
In each example, it is assumed that the number of breakpoints of the transmission function is known.
To avoid any inverse crime in the numerical experiments, the simulated data in the forward model is generated via a method (brute force numerical integration) different from the method used for the inversion (based on an analytic formula).

The paper is organized as follows.  In Section \ref{section:ForwardModel}, we review the forward model of coherence propagation and the closed-form solution described in \cite{Beckus:17}.  In Section \ref{section:InverseProblem}, we formulate the inverse problem and describe the gradient-descent-based algorithm. The algorithm is demonstrated with simulated data in Section \ref{section:Results} and with experimental measurements in Section \ref{section:ResultsExperiment}.  In Section \ref{section:Discussion}, we discuss possible extensions to this work. The derivation of the analytic gradients used by the algorithm are detailed in Appendix \ref{section:DeriveGradients}.

\section{Background: Forward Model of Coherence Propagation}
\label{section:ForwardModel}
\begin{figure}[t!]
\centering
\includegraphics[scale=1]{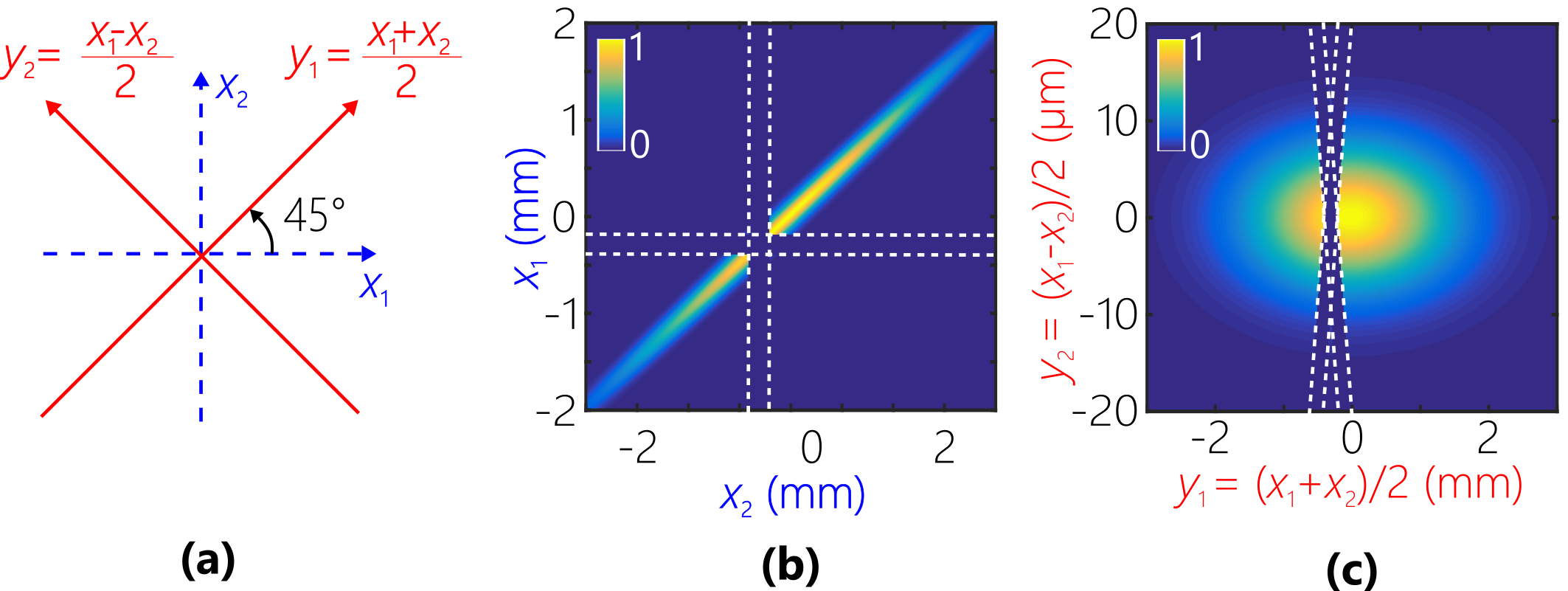}
\caption{(a) Illustration of rotated coordinates. An example of a generalized source is shown in (b) unrotated coordinates and (c) rotated coordinates.  For this example, the Gauss-Schell source parameters are $A\!=\!1$, $w\!=\!1$~mm, $\sigma\!=\!50~\mu$m.  The transmission function is such that
$t(x)\!\!=\!\!0$ for $x \!\in\! [a_1,a_2)$, and $t(x) \!\!=\!\! 1$ otherwise, where $a_1\!=\!-0.4$~mm and $a_2\!=\!-0.2$~mm. Dotted white lines indicate the regions affected by the transmission function.
Reprinted from \cite{Beckus:17}.
}
\label{Fig1:Overview}
\end{figure}

 Given a realization $U(x)$ of a random field, the coherence function is defined as the two-point correlation  $G(x_1,x_2)\!=\!\langle U(x_{1})U^{*}(x_{2})\rangle$, with $\langle\cdot\rangle$ representing the ensemble average  \cite{BornWolf:99}.  We will work with the coherence function in rotated coordinates
\begin{align} \label{eqn:rotatedcoordinates}
y_1=\frac{x_1+x_2}{2}, \qquad y_2=\frac{x_1-x_2}{2},
\end{align}
as illustrated in Fig.~\ref{Fig1:Overview}.  We refer to the coordinate $y_1$ as the {\em intensity} coordinate, while to $y_2$ as the {\em coherence} coordinate.

We will work in the Fresnel regime where the propagation of the coherence is given by
\begin{align} \label{fresneliny}
G_d(y_1,y_2)=
\frac{k}{\pi d}
\iint_{\mathbb{R}^2} G(y_1', y_2') \, \mathcal{L}(y_1,y_1',y_2,y_2') \, dy_1' \, dy_2'
\end{align}
with kernel
\begin{align}
\mathcal{L}(y_1,y_1',y_2,y_2') = \exp \left\{i\frac{2k}{d}(y_1-y_1')(y_2-y_2') \right\}\:,
\end{align}
where $d$ is the propagation distance and $k$ is the wavenumber.
The integration is over $\mathbb{R}^2$, i.e. the infinite plane of the source.

Let $\Gaussian{\beta}{x}\!=\!\exp\{-x^{2}/2\beta^{2}\}$ denote the Gaussian of standard deviation $\beta$.
A quasi-homogeneous Gauss-Schell beam
\begin{align}\label{eqn:GaussSchell}
G^-(y_{1}',y_{2}')=A \, \exp\{iy_{1}'y_{2}'/R^{2}\} & \Gaussian{w}{y_{1}'} \, \Gaussian{\sigma}{y_{2}'},
\end{align}
(where $A$ is the amplitude, $w$ the width of the intensity profile, $\sigma$ the coherence width, and $R$ the radius of curvature of an acquired quadrature phase) 
upon propagation over a distance $d$, results in another Gauss-Schell beam
\begin{align}\label{eqn:GaussSchellProp}
\widetilde{G}_d(y_1,y_2)
= \tA \, \exp\big(iy_{1}y_{2}/\tR^{2}\big) \Gaussian{\tw}{y_{1}} \, \Gaussian{\tsigma}{y_{2}}
\end{align}
with transformed parameters
\begin{subequations} \label{eqn:doublepropparameters}
\begin{align}
\tilde{A} &= \frac{A}{(1+\delta)\sqrt{1+\xi^{2}}}, \\
\tilde{R} &= R\,\sqrt{\frac{(1+\delta)(1+\xi^2)}{1+(1+\frac{1}{\delta})\xi^2}}, \\
\tilde{w} &= w\,(1+\delta)\sqrt{1+\xi^{2}},\\
\tilde{\sigma} &= \sigma\,(1+\delta)\sqrt{1+\xi^{2}},
\end{align}
\end{subequations}
where $\lambda$ is the wavelength, $\xi\!=\!\ell^{2}/\{w \sigma(1+\delta)\}\!=\!d/z_{\mathrm{GS}}$, $z_{\mathrm{GS}}\!=\!4\pi\sigma w(1+\delta)/\lambda$ is a scaled Rayleigh range, $\ell\!=\!\sqrt{d/2k}$, and $\delta\!=\!\ell^{2}/R^{2}$ \cite{FRIBERG1982383,GORI1983149}.

A generalized source is defined as a Gauss-Schell beam \eqref{eqn:GaussSchell} modulated by a piecewise constant transmission function $t$:
\begin{align} \label{eqn:generalized_source}
G(y_1',y_2')
=& G^-(y_{1}',y_{2}') t(y_1'+y_2') \, t^*(y_1'-y_2'),
\end{align}
which follows from the definition of coherence, and the transformations found in \eqref{eqn:rotatedcoordinates}.
For $N$ arbitrarily fixed, $-\infty=a_0 \!<\! a_1 \!<\! \cdots \!<\! a_{N} \!<\! a_{N+1}=\infty$, $t(x)=c_j$ for $x\in [a_j,a_{j+1})$, where each $c_j$ is a complex-valued constant, $j=0,\ldots,N$. 

The main result of \cite{Beckus:17}, restated below, provides an approximation to the propagated coherence function \eqref{fresneliny} for a generalized source  characterized in terms of the coherence of the Gauss-Schell field propagated in free space and a multiplicative term capturing the modification due to interaction with the transmission function. This approximation has a closed form in terms of a conjugated Hilbert transform
\begin{align}\label{eqn:ConjugatedHilbertTransform}
H^u f(\omega):=\exp(-i\omega u) \, \mathrm{p.v.}{\frac{1}{\pi}}\int\frac{\exp(isu)\,f(s)}{\omega-s}ds\:,
\end{align}
where $\mathrm{p.v.}$ stands for principal value.
\begin{thmnonum}[\cite{Beckus:17}]\label{main_result}
A generalized source as in \eqref{eqn:generalized_source} satisfying
\begin{subequations}\label{eqn:tRestrictions}
\begin{align}
w > 10^2 \sigma > 10^3 \lambda&, \label{eqn:AlphaSigmaRegime}
\\
\sum_{j=1}^{N}  \Gaussian{w}{|a_{j}|-3\sigma} &< 4, \label{eqn:tRestriction1}
\\
\min_{j=2,\ldots, N}(a_{j}-a_{j-1}) &> 3\sigma, \label{eqn:PartitionSize}
\end{align}
\end{subequations}
is situated at the plane $z\!=\!0$.  At the detection plane $z\!=\!d$, the coherence $G_d(y_1,y_2)$ is well approximated by
\begin{align}\label{eqn:mainresult1}
\Gdapprox(y_1,y_2)
=& \widetilde{G}_d(y_1,y_2) \frac{i}{2 \Gaussian{\eta \widetilde{\sigma}}{y_2}} \nonumber \\
&\times \sum_{j=2}^{N} T_{j,j} \left[ \left(H^{b_{j}(y_1)} - H^{b_{j-1}(y_1)} \right)\GaussianNoParm{\tsigma / \eta}\right](y_2)
\end{align}
where 
$T_{j,j} = |t(x)|^2$ for $x \in [a_{j-1},a_j)$, and
\begin{subequations}\label{eqn:mainresult_vars}
\begin{align}
\eta &= \sqrt{1+\frac{\sigma^2 \tsigma^2}{\ell^4}}, 
\\
b_j(y_1) &= \frac{1}{\eta^2 \ell^2} \left( a_j - \frac{y_1}{(1+\delta)(1+\xi^2)} \right).\label{bjs}
\end{align}
\end{subequations}
\end{thmnonum}
In \eqref{eqn:mainresult1}, the conjugated Hilbert transform for two different parameters is applied to the specified Gaussian according to \eqref{eqn:ConjugatedHilbertTransform}.
The hypotheses in \eqref{eqn:tRestrictions} are satisfied in Section \ref{section:Results} below.  However, these conditions are merely sufficient for \eqref{eqn:mainresult1} to hold.  This is demonstrated in Section \ref{section:ResultsExperiment}, where the experimental parameters violate the first inequality of \eqref{eqn:AlphaSigmaRegime}, yet the approximation is still dependable and allows for successful inversion. 
Further details on the physical meaning and the analysis of the result above can be found in \cite{Beckus:17}.

For a Gauss-Schell source, truncation of the transmission function away from the mean (e.g.,  at $|y_1'| = 3w$) is  insignificant to the approximation, allowing us to set  $T_{1,1}=T_{N+1,N+1}=0$.

\section{A minimum residual approach to the Inverse Problem} \label{section:InverseProblem}

Using a set of measured coherence samples, we seek to determine the breakpoints $\avec = (a_1,\cdots,a_N)$ of a generalized source, as well as the distance $d$ between the source and the measurement plane.

For a trial vector $\avec = (a_1,\cdots,a_N)$ and some $d>0$, we consider the residual between the measured coherence $G_d$ and the approximation $\Gdapprox$ calculated using \eqref{eqn:mainresult1}:
\begin{align}\label{eqn:pointerror}
\Gerror(y_1,y_2;\avec,d) = \Gdapprox(y_1,y_2;\avec,d) - G_d(y_1,y_2)
\end{align}for each pair of measurements $(y_1,y_2)$. More precisely, given the sample points $(\sampleyI{k},\sampleyII{k})$, $k=1,\ldots,\SampleCount$, we introduce the objective function
\begin{align} \label{eqn:ErrorAllSamples}
\GerrorSquaredSum(\avec,d)
= \frac{1}{\SampleCount} \sum_{k=1}^{\SampleCount} |\Gerror(y_1^k,y_2^k; \avec,d)|^2.
\end{align}

We consider the problem of minimizing $\GerrorSquaredSum$ with respect to the parameters $\avec,d$, using a gradient-descent algorithm \cite{ChongZak:99}.
The fixed-size steps are described by
\begin{align}
\aveciter{n+1} &= \aveciter{n} - \StepSizBreakpoint \,\, \dGerrorSquaredSumda,
\\
\diter{n+1} &= \diter{n} -  \StepSizDistance \,\, \frac{\partial \GerrorSquaredSum}{\partial d},
\end{align}
where $n$ is the gradient-descent iteration number.
Of novelty here, when a local minimum has been found, i.e., when the partial derivatives $\frac{\partial \GerrorSquaredSum}{\partial \avec}, \frac{\partial \GerrorSquaredSum}{\partial d}$ both fall below prescribed thresholds $\ThresholdGradBreakpoint,\ThresholdGradDistance$,
the algorithm performs an additional check for a global minimum.  This is accomplished by verifying that the residual is insignificant at \textit{each sample point}, specifically
\begin{align} \label{eqn:GlobalMinCheck}
|\Gerror(\sampleyI{k},\sampleyII{k})| < \ThresholdError, \quad 1 \le k \le M.
\end{align}

As will be seen in Fig. \ref{Fig3:OneObjectDescent}(b) of the first example, a characteristic of the global minimizer is that the actual and estimated coherence functions closely match at all sample points, and thus the residual is small at each point.
If condition \eqref{eqn:GlobalMinCheck} is not met, then the algorithm is randomly re-initialized with a starting point in the admissible domain.

The partial derivatives of $\GerrorSquaredSum$ with respect to the breakpoints admit an analytic closed form as follows.  Let $\realpart{.}$ and $\imagpart{.}$ denote the real and imaginary components of their complex argument, respectively.  Then,
\begin{align} \label{eqn:gradient}
\frac{\partial \GerrorSquaredSum}{\partial a_j} 
=& \sqrt{\frac{2}{\pi}} \frac{\tA \tsigma}{\eta^3 \ell^2 \SampleCount}
\left( T_{j,j} - T_{j+1,j+1} \right)
\nonumber \\
\times& \sum_{k=1}^{\SampleCount} \left\{ \realpart{\Gerror(\sampleyI{k},\sampleyII{k})} \, \realpart{\GradMain{j}(\sampleyI{k},\sampleyII{k})}
\right.
\nonumber \\
& \left. - \imagpart{\Gerror(\sampleyI{k},\sampleyII{k})} \, \imagpart{\GradMain{j}(\sampleyI{k},\sampleyII{k})}  \right\},
\end{align}
where
\begin{subequations}       
\begin{align} \label{eqn:gradmain}
\GradMain{j}(y_1,y_2)
&= \GradGauss{j}(y_1,y_2) \exp \left\{ i y_2 b_{j}(y_1) - i y_{1}y_{2}/\tR^{2} \right\},
\\
\GradGauss{j}(y_1,y_2)
&= \Gaussian{\tw}{y_1} \Gaussian{ \frac{\ell^2 \eta}{\sigma} }{y_2} \Gaussian{\eta / \tsigma}{b_j(y_1)};
\end{align}
\end{subequations}see Appendix \ref{section:DeriveGradients}.

The derivative can also be calculated for measurements of the degree of spatial coherence
$g_d(y_1,y_2) = G_d(y_1,y_2)/\sqrt{I_1 I_2}$, where $I_1=G_d(y_1+y_2,0)$ and $I_2=G_d(y_1-y_2,0)$ are the intensities at the first and second correlation points.  The approximated degree of coherence is likewise defined as $\gdapprox(y_1,y_2) = \Gdapprox(y_1,y_2)/\sqrt{\IdapproxI \IdapproxII}$, where $\IdapproxI,\IdapproxII$ are the corresponding approximated intensities calculated using \eqref{eqn:mainresult1}.  We will denote the complex conjugate of $\gdapprox$ by $\gdapprox^*$. In this case, \eqref{eqn:gradient} still holds with transformations $\Gerror \!\rightarrow\! \gerror$ and $\GradMain{j} \!\rightarrow\! \GradMainNorm{j}$ where
\begin{subequations}
\begin{align}\label{eqn:pointerror_normalized}
\gerror(y_1,y_2;\avec,d) =& \gdapprox(y_1,y_2;\avec,d) - g_d(y_1,y_2),
\\
\GradMainNorm{j}(y_1,y_2)
=& \frac{1}{\sqrt{\IdapproxI \IdapproxII}}
\left\{ \GradMain{j}(y_1,y_2) - \gdapprox^*(y_1,y_2)
\right.
\nonumber \\
&\left. \times \left[ \IdapproxI Z_j(y_1-y_2,0) + \IdapproxII Z_j(y_1+y_2,0) \right]
\right\}. \label{eqn:gradmain_normalized}
\end{align}
\end{subequations}

The derivative with respect to the distance is calculated by a finite difference.

\section{Applications with simulated data} \label{section:Results}

\subsection{Single object at known distance} \label{section:ResultsOneObject}

\begin{figure}[t!]
\centering
\includegraphics[scale=1]{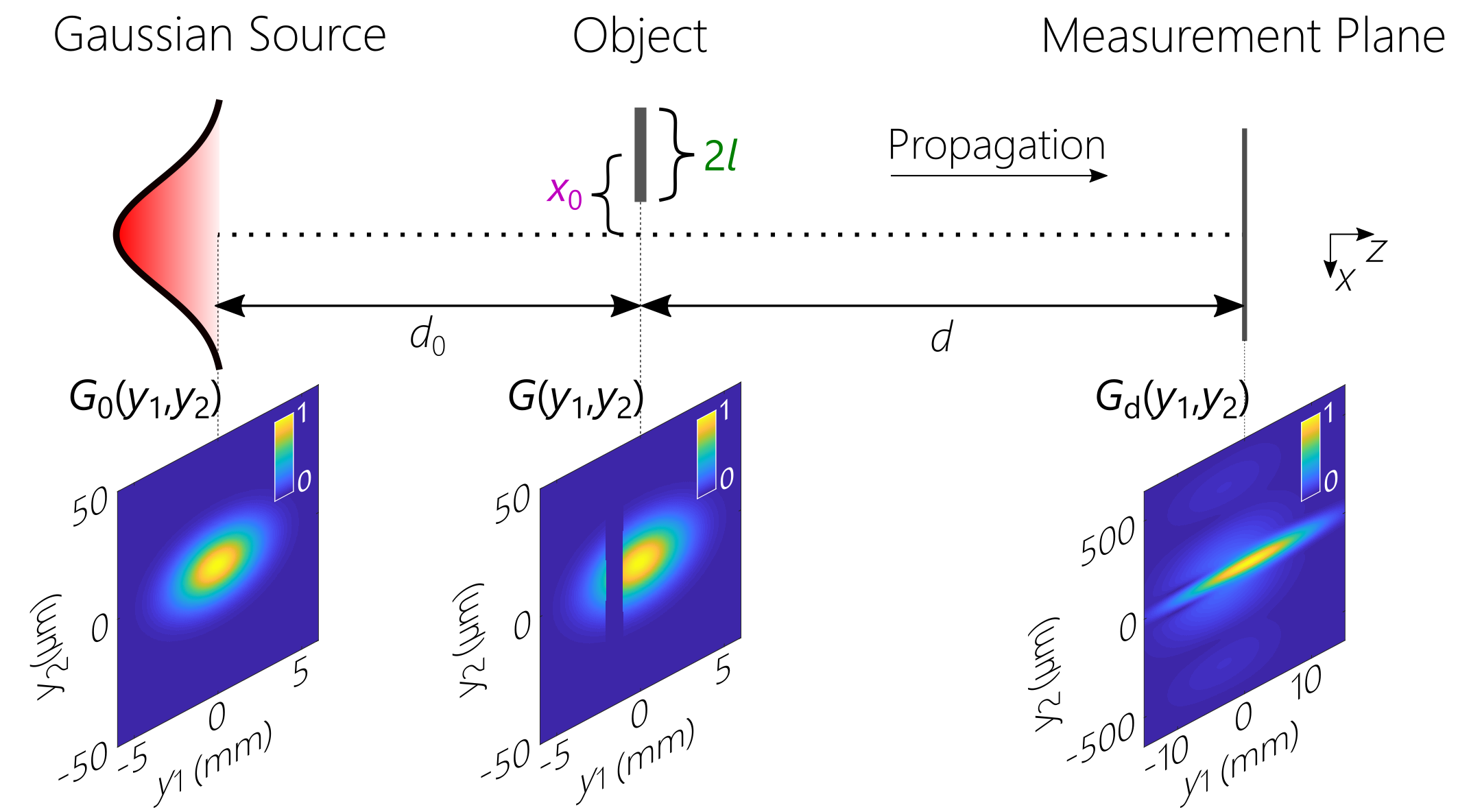}
\caption{A single object scene 
with $x_0\!=\!-1.5$~mm, $l\!=\!0.5$~mm, $d_0\!=\!10$~cm, and $d\!=\!100$~cm.  The normalized magnitude of the coherence function is shown at the bottom of the diagram in three planes: in the plane of the Gaussian source, immediately after interacting with the object (i.e. at the secondary source), and at the measurement plane.
}
\label{Fig2:OneObjectScene}
\end{figure}

Consider a Gauss-Schell source at $z=0$ propagating a distance $d_0$ in free space, where it is blocked by a single object of width $2l$ centered along the transverse axis at the offset point $x\!=\!x_0$ as shown in Fig.~\ref{Fig2:OneObjectScene}. The detector is located at a distance $d$ from the object plane. 
In this first example we seek to estimate the parameters $x_0$ and $l$, assuming that the distances $d_0$ and $d$ are known.

The parameters for the original Gaussian source (at $z=0$) are amplitude $A\!=\!1$, width $w\!\approx\!1.7$~mm (yielding an intensity full width at half-maximum (FWHM) of 4 mm), and variance $\sigma\!\approx\!8.5~\mu$m (yielding a coherence FWHM of 20 $\mu$m). Also we assume the source has no phase (i.e., in the limit as $R \rightarrow \infty$). The wavelength is $\lambda\!=\!633$~nm.

In the forward model, the parameters $\tilde{A}$, $\tilde{R}$, $\tilde{w}$, and $\tilde{\sigma}$ in the plane of the object are calculated using the transformations in \eqref{eqn:doublepropparameters}.
The object is modeled using \eqref{eqn:generalized_source} with 
$N\!=\!2$, and the breakpoints $a_1 \!=\! x_0\!-\!l$, and $a_2 \!=\! x_0\!+\!l$, and the coherence is propagated from the object plane to the detector plane using \eqref{fresneliny} to obtain the coherence measurements.
In solving the inverse problem, the estimated coherence is calculated by \eqref{eqn:mainresult1}.

The initial offset location parameter $x_0$ is set to uniformly span an admissible domain, whose bounds ($\pm 7.1$mm) are dependent on the width of the source Gaussian. The initial length $l$ is assigned between $0$ and $2$~mm at random. The other parameters are fixed, $\StepSizBreakpoint\!=\!10^{-4}$, $\ThresholdGradBreakpoint\!=\!10^{-2}$, and
$\ThresholdError\!=\!2 \times 10^{-3}$.

\begin{figure}[t!]
\centering
\includegraphics[scale=1]{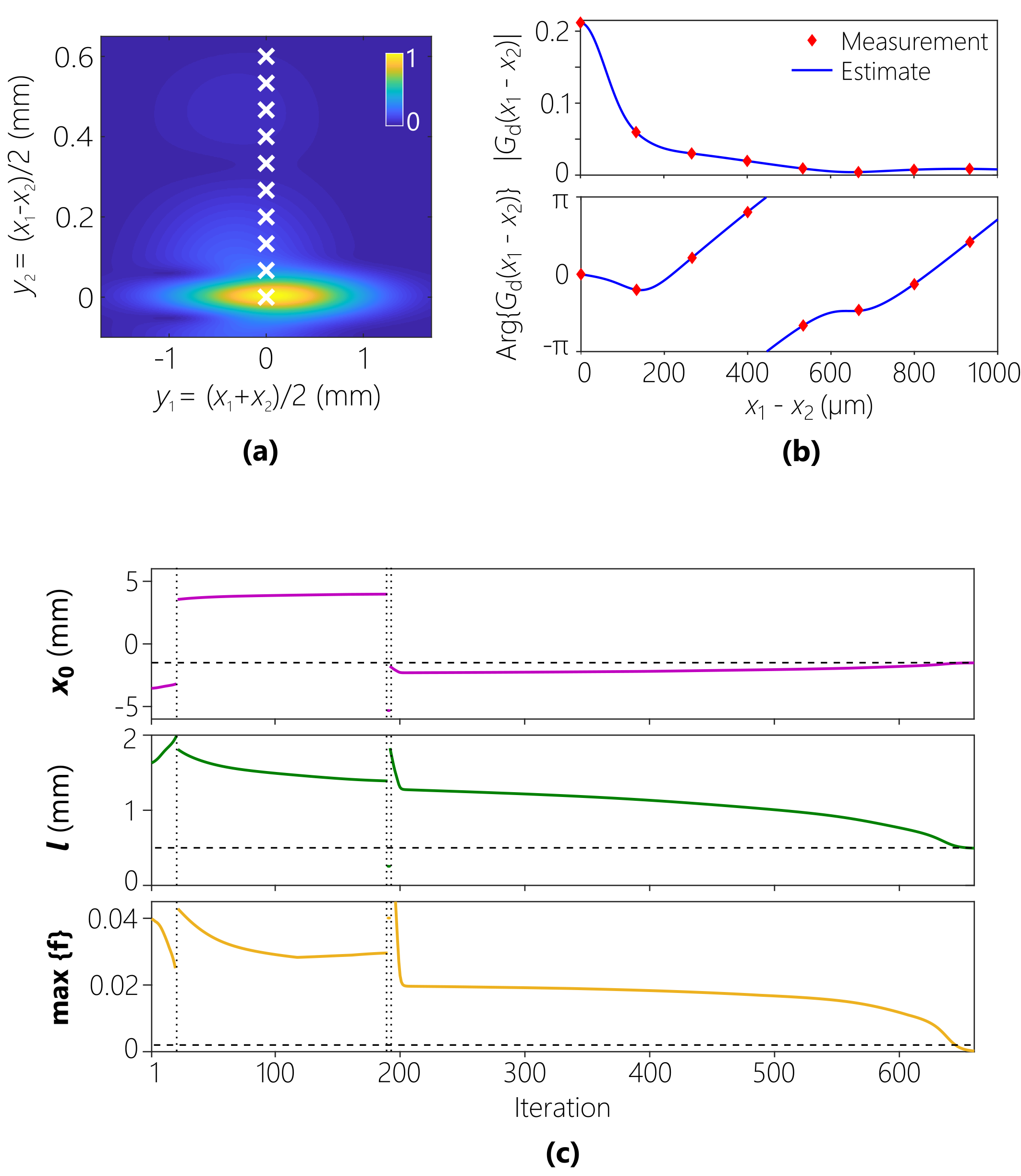}
\caption{Reconstruction results for one object at known distance.
(a) Normalized modulus of coherence function in the measurement plane with sample points marked.
(b) Modulus and phase of coherence function at measurement plane. Both measured samples and final estimate are shown.
(c) Path of the gradient descent algorithm.  The top two plots show the estimates of the two parameters, with horizontal dashed lines indicating the actual value of the parameters.  The bottom plot shows the maximum residual value among all sample points with the threshold $\ThresholdError$ indicated by a dashed line.  A vertical dotted line indicates a restart of the algorithm with a new initialization.}
\label{Fig3:OneObjectDescent}
\end{figure}

The results of one execution of the gradient-descent algorithm are shown in Fig.~\ref{Fig3:OneObjectDescent}.  The actual parameter values are $x_0\!=\!-1.5$~mm, $l\!=\!500$~$\mu$m, and $d\!=\!1$~m.  The modulus of the actual simulated coherence function is shown in Fig.~\ref{Fig3:OneObjectDescent}(a), with the sample points marked.  The measured coherence function at the 10 sample points is shown in Fig.~\ref{Fig3:OneObjectDescent}(b) along with the final estimate (calculated using \eqref{eqn:mainresult1}).  The dynamics (with iterations) are displayed in Fig.~\ref{Fig3:OneObjectDescent}(c).  The parameter estimates are shown in the top two plots, with the actual value indicated by horizontal dashed lines.  The maximum residue, defined as $\MaxGerror := \max_{y_2} f(0,y_2)$ is shown in the bottom plot with the threshold $\ThresholdError$ indicated by a horizontal dashed line.
Vertical dotted lines indicate where a new initialization point is chosen and the algorithm restarted.
This restart can be triggered when the partial derivatives fall below the threshold $\ThresholdGradBreakpoint$ while $\MaxGerror > \ThresholdError$, indicating that the local minimum is not a global minimum.
The restart may also be triggered when the parameters leave the admissible domains.
In the final iterations, it can be seen that the parameter estimates converge to the correct values and $\MaxGerror$ falls below the threshold.
The small residue is evidenced by the excellent agreement between the measured and estimated coherence functions in Fig.~\ref{Fig3:OneObjectDescent}(b).   The estimates are $x_0\!=\!-1.504$~mm and $l\!=\!497.2$~$\mu$m, which have an error of less than 1\% (an error which could be made arbitrarily small by reducing the value of $\ThresholdGradBreakpoint$).

A video, one frame of which is displayed in Fig.~\ref{Fig4:OneObjectVideo}, is also provided in supplementary material showing the progression of the algorithm.
\begin{figure}[t!]
\centering
\includegraphics[scale=1]{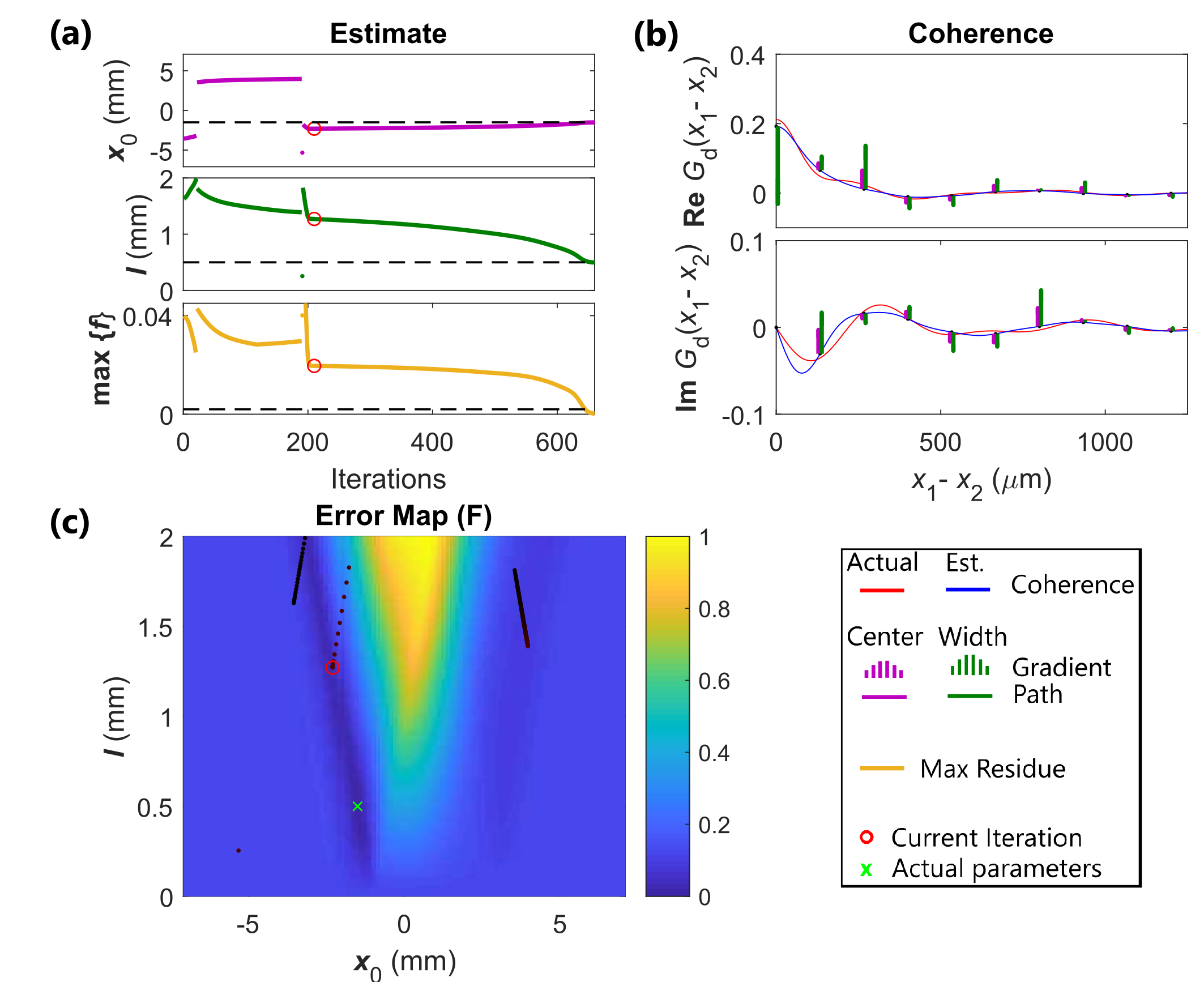}
\caption{Video showing gradient descent for single object (see Visualization 1 in supplementary material).
The plots in (a) are the same as in Fig. \ref{Fig3:OneObjectDescent}(c), and are included in the video to show the progression of the algorithm in each frame.
The real and imaginary parts of the coherence functions are shown in (b).  Bars extending vertically from a sample point indicate the magnitude and direction of the step contribution from that point (up indicates the parameter increases at the next step).  The algorithm calculates the next step by accumulating the individual contributions from each point.
(c) contains a map of residual function $\GerrorSquaredSum$ from \eqref{eqn:ErrorAllSamples} plotted with regard to the two object parameters.
The location of the actual parameters is marked by a green ``x''.  The estimate at the current iteration is indicated with a red circle in (a) and (c).
}
\label{Fig4:OneObjectVideo}
\end{figure}
Fig.~\ref{Fig4:OneObjectVideo}(a) corresponds to Fig.~\ref{Fig3:OneObjectDescent}(c) and shows the path of the iterations of the $x_0$ and $l$ estimates, as well as $\MaxGerror$.  Fig.~\ref{Fig4:OneObjectVideo}(b) corresponds to Fig.~\ref{Fig3:OneObjectDescent}(b) and displays the actual coherence function, as well as the approximation based on the current parameter estimates. The values of the partial derivatives of $\Gerror$ are indicated as bars extending vertically from the sample points.  The residual map $\GerrorSquaredSum$ is displayed in Fig.~\ref{Fig4:OneObjectVideo}(c) as a function of the two parameters $x_0$ and $l$, and the path of the estimates is indicated in the map.

\subsection{Single object at unknown distance} \label{section:ResultsOneObjectDistanceUnknown}

\begin{figure}[t!]
\centering
\includegraphics[scale=1]{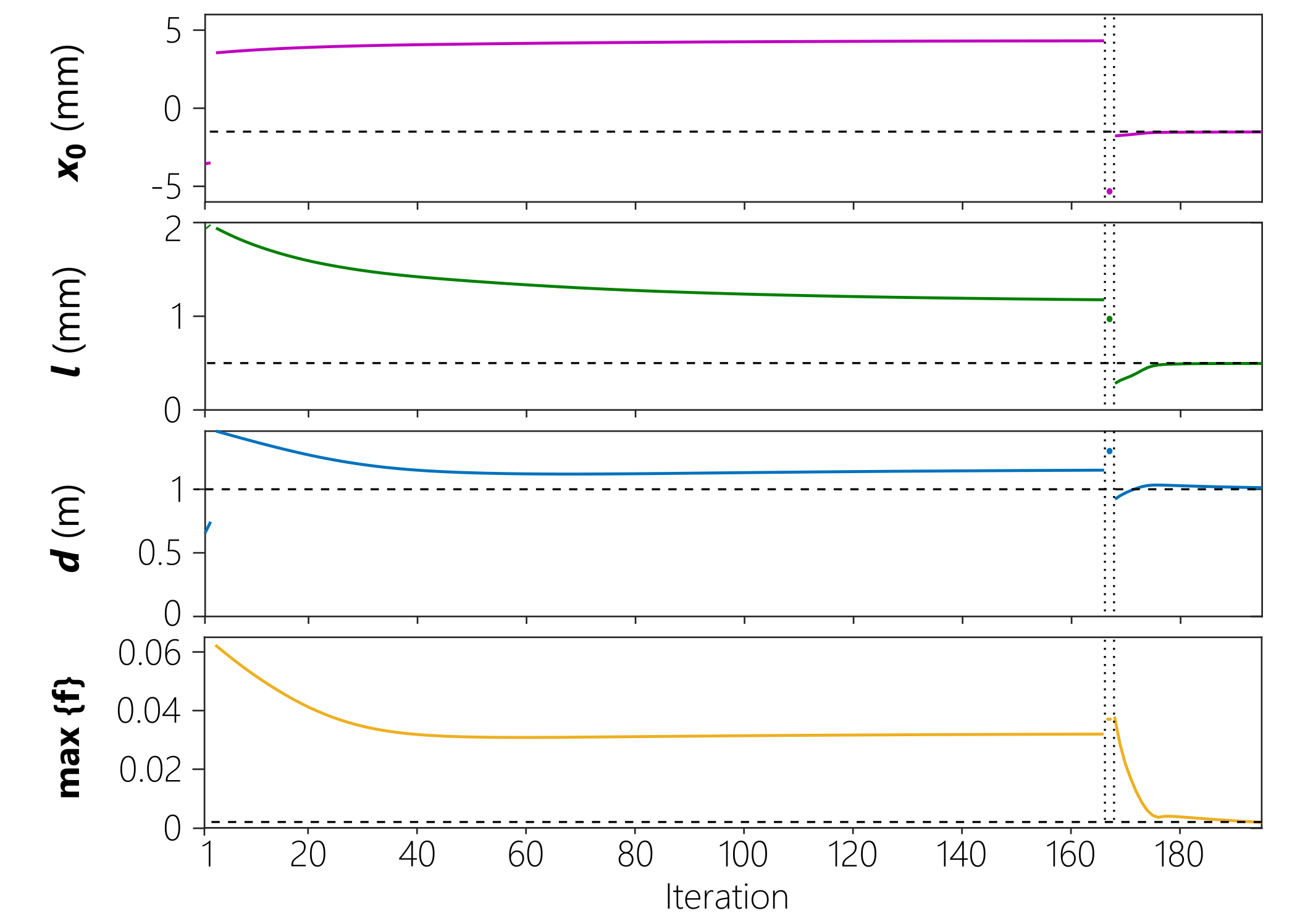}
\caption{Gradient descent algorithm estimating three object parameters: $x_0$, $l$, and $d$.  The configuration and sample points are the same as in  Fig. \ref{Fig3:OneObjectDescent}.  The plot labels are the same as those defined in Fig. \ref{Fig3:OneObjectDescent}(c), with an additional plot included for parameter $d$.
}
\label{Fig5:OneObjectDescent3d}
\end{figure}

We now expand on the previous example by estimating a third parameter, the distance $d$ between the object and measurement plane.  The partial derivative of the residual $\GerrorSquaredSum$ with respect to distance is calculated by a finite difference, with $\ThresholdGradDistance=10^{-3}$.  When a new initialization point is generated, the distance $d$ is randomly assigned from an admissible domain between 0.5~m and 1.5~m.
The results of the algorithm using simulated measurements are shown in Fig.~\ref{Fig5:OneObjectDescent3d}.
The actual parameters are the same as in the previous section, and the estimated values are $x_0\!=\!-1.521$~mm, $l\!=\!496.0$~$\mu$m, and $d\!=\!1.013$~m.
As with the two-parameter example, the estimate is close with a maximum parameter error of less than 1.5\% (and could be reduced by using smaller gradient thresholds).

\subsection{Two intercepting objects}

\begin{figure}[t!]
\centering
\includegraphics[scale=1]{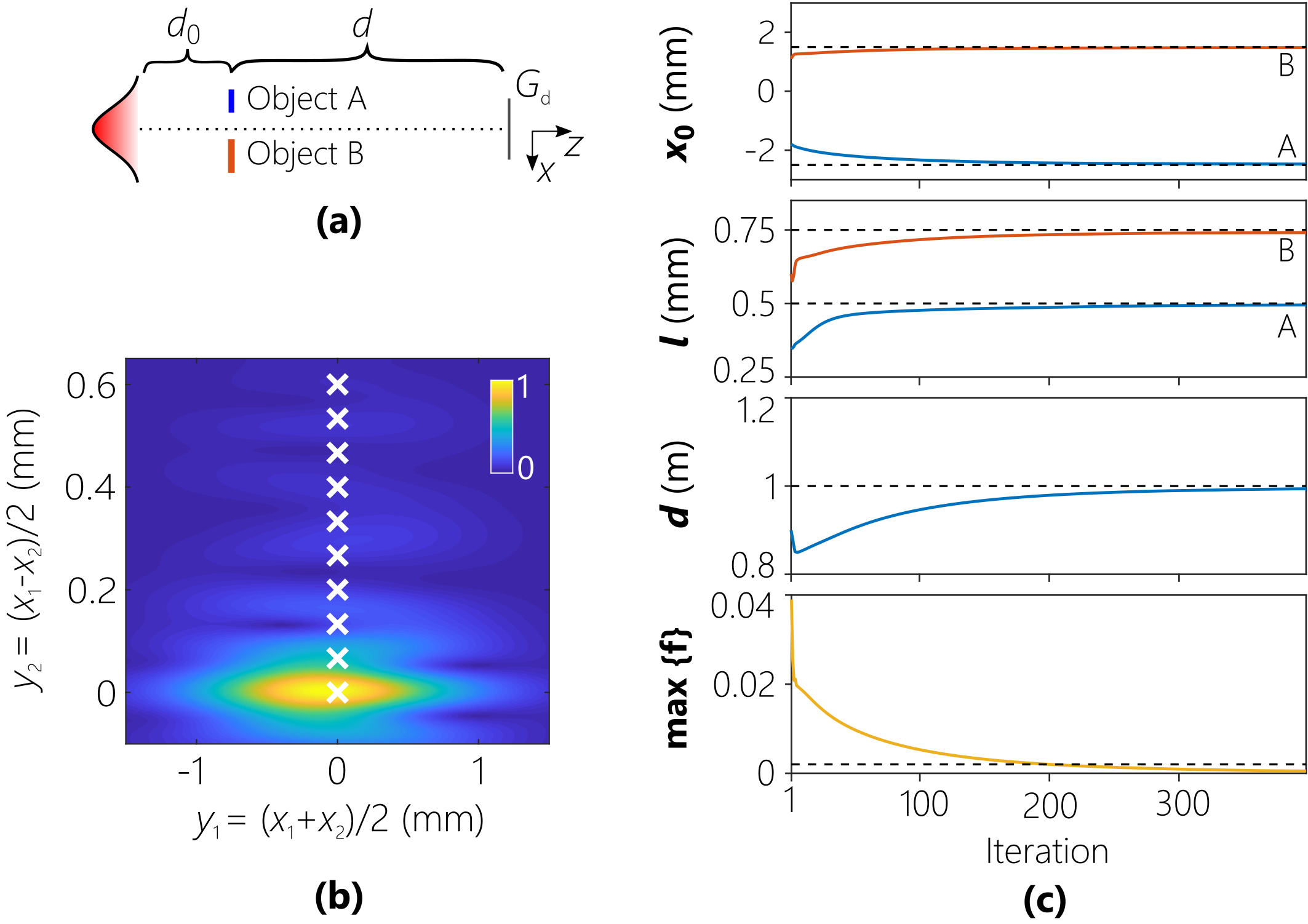}
\caption{Example showing estimation of positions of two objects in the same axial plane.
(a) Diagram of scenario.
(b) Normalized modulus of coherence function in the measurement plane with sample points marked.
(c) Path of the gradient descent algorithm.  The top three plots show the estimates of the five parameters (blue lines correspond to Object A and orange lines to Object B), with dashed lines indicating the actual value of the parameters.  The bottom plot shows the maximum residual value among all sample points with the threshold $\ThresholdError$ indicated by a dashed line.
The Gaussian source parameters are the same as in the one object example.  The object parameters are $\TwoObjectParam{x_0}{A}\!=\!-2.5$~mm, $\TwoObjectParam{l}{A}\!=\!500$~$\mu$m for Object A, $\TwoObjectParam{x_0}{B}\!=\!1.5$~mm and $\TwoObjectParam{l}{B}\!=\!750$~$\mu$m for Object B.  The two objects are located in the same plane, and the actual distances are $d_0\!=\!0.1$~m, $d\!=\!1$~m.
The final estimates are $\TwoObjectParam{x_0}{A}\!=\!-2.483$~mm, $\TwoObjectParam{l}{A}\!=\!495.6$~$\mu$m, $\TwoObjectParam{x_0}{B}\!=\!1.492$~mm, $\TwoObjectParam{l}{B}\!=\!741.3$~$\mu$m, and $d\!=\!0.944$~m.
}
\label{Fig6:TwoObjectSamePlane}
\end{figure}

We now demonstrate the ability of the algorithm to handle more complicated scenes with more parameters.  Fig.~\ref{Fig6:TwoObjectSamePlane} shows the results for a five-parameter estimation problem in which two objects are located in the same plane. The parameters are the center $\TwoObjectParam{x_0}{A}$ and half-width $\TwoObjectParam{l}{A}$ of the first object defined by breakpoints $a_1$ and $a_2$, the center and half-width parameters for the second object ($\TwoObjectParam{x_0}{B}$ and $\TwoObjectParam{l}{B}$) defined by breakpoints $a_3$ and $a_4$, and the distance $d$ between the object and measurement planes.  The algorithm parameters $\StepSizBreakpoint$, $\ThresholdGradBreakpoint$, $\ThresholdError$, and $\ThresholdGradDistance$ are the same as in Sections \ref{section:Results}.\ref{section:ResultsOneObject} and \ref{section:Results}.\ref{section:ResultsOneObjectDistanceUnknown}, and we use the same approach as with the one object example, only with two additional breakpoints.
The coherence function and sample points are shown in Fig.~\ref{Fig6:TwoObjectSamePlane}(b).  The iterations are shown in Fig. \ref{Fig6:TwoObjectSamePlane}(c).

The maximum error in parameter estimates is less than $1.2\%$; very small considering that only 10 sample points were used along the coherence axis.

\begin{figure}[t!]
\centering
\includegraphics[scale=1]{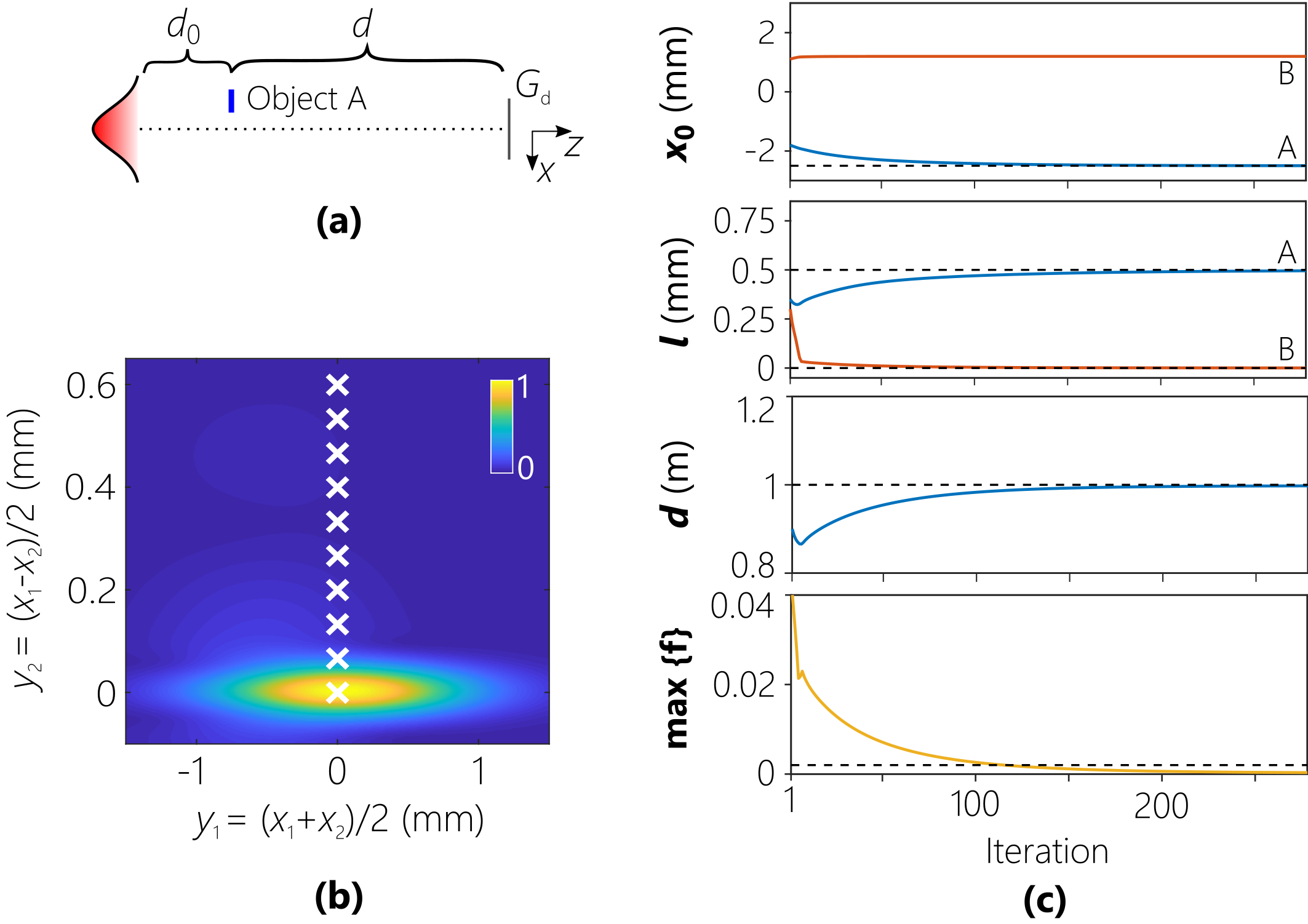}
\caption{
Example showing estimation of positions assuming two objects in the same axial plane when only one object is actually present.
Panels (a)-(c), as well as the source parameters, are the same as in Fig.~\ref{Fig6:TwoObjectSamePlane}.  The parameters for Object A are $\TwoObjectParam{x_0}{A}\!=\!-2.5$~mm, $\TwoObjectParam{l}{A}\!=\!500$~$\mu$m, and Object B is absent from the scene.  The actual distances are $d_0\!=\!0.1$~m, $d\!=\!1$~m.
The final estimates are $\TwoObjectParam{x_0}{A}\!=\!-2.494$~mm, $\TwoObjectParam{l}{A}\!=\!495.5$~$\mu$m, $\TwoObjectParam{x_0}{B}\!=\!1.198$~mm, $\TwoObjectParam{l}{B}\!=\!0.45$~$\mu$m, and $d\!=\!0.998$~m.  Note that the estimate of $\TwoObjectParam{l}{B} \approx 0$, indicating no Object B is present (thus rendering the estimate of $\TwoObjectParam{x_0}{B}$ irrelevant). }
\label{Fig7:TwoObjectSamePlaneOverestimate}
\end{figure}

While in the previous examples we have assumed the number of objects is known, it is also possible to use the algorithm when all we have is a crude upper bound on the number of objects.  Additional breakpoints can be included in the transmission function, and the ``missing'' objects will be estimated as having zero width.  To illustrate this point, we repeat the previous example of Fig.~\ref{Fig6:TwoObjectSamePlane}, but with Object B removed.  The results are shown in Fig.~\ref{Fig7:TwoObjectSamePlaneOverestimate}.  The parameters of Object A are correctly estimated, whereas because the second assumed object is not actually present, the estimated width of Object B rapidly approaches zero.

\begin{figure}[t!]
\centering
\includegraphics[scale=1]{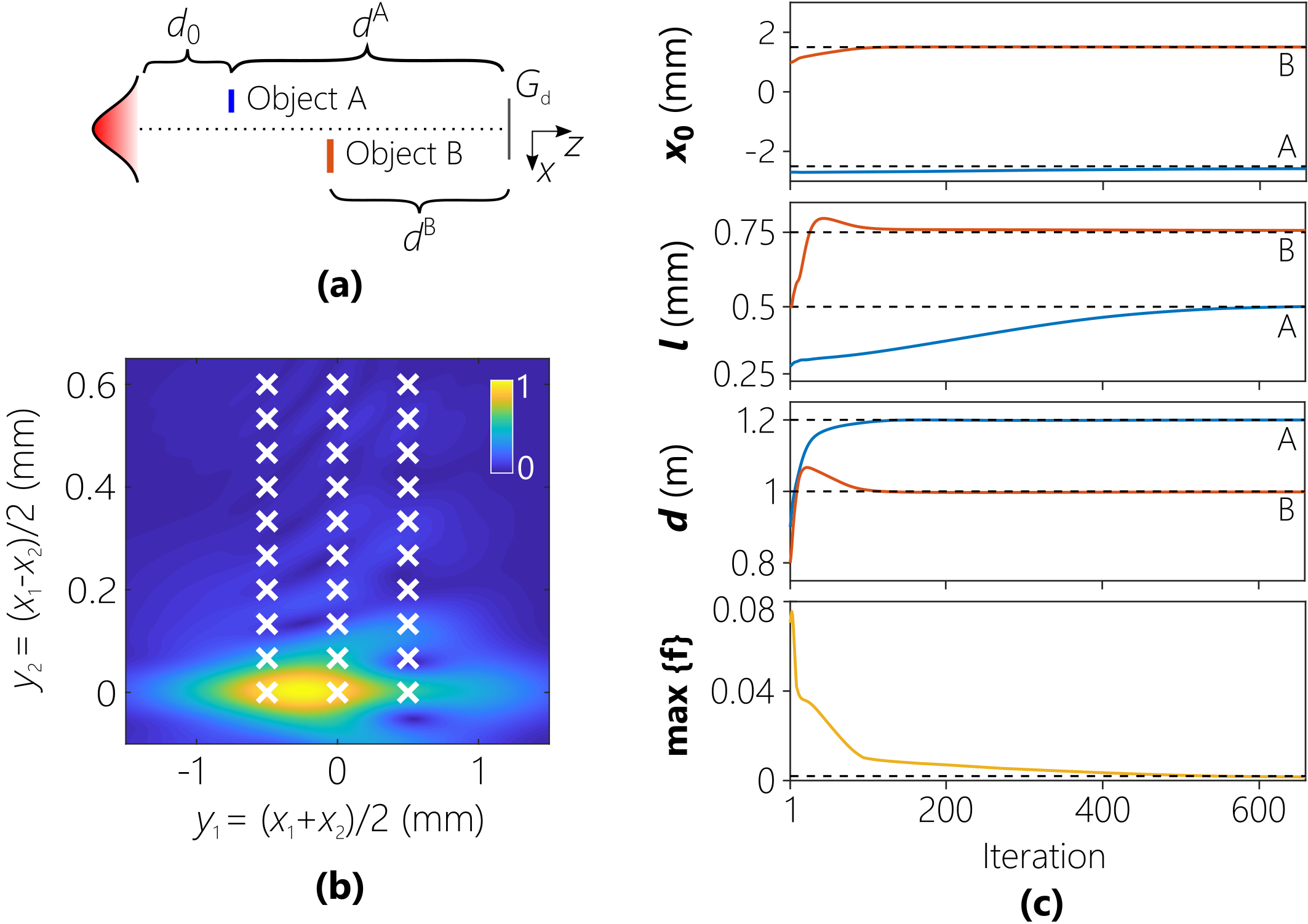}
\caption{
Example showing estimation of positions of two objects in different axial planes.
Panels (a)-(c) are the same as in Fig.~\ref{Fig6:TwoObjectSamePlane}.
The Gaussian source and object parameters are also the same as used in Fig.~\ref{Fig6:TwoObjectSamePlane}.  The distances are $d_0\!=\!0.1$~m, $\TwoObjectParam{d}{A}\!=\!1.2$~m, and $\TwoObjectParam{d}{B}\!=\!1$~m.
The final estimates are $\TwoObjectParam{x_0}{A}\!=\!-2.516$~mm, $\TwoObjectParam{l}{A}\!=\!501.1$~$\mu$m, $\TwoObjectParam{x_0}{B}\!=\!1.508$~mm, $\TwoObjectParam{l}{B}\!=\!755.7$~$\mu$m, $\TwoObjectParam{d}{A}\!=\!1.20$~m, and $\TwoObjectParam{d}{B}\!=\!0.998$~m.
}
\label{Fig8:TwoObjectDiffPlanes}
\end{figure}

Fig.~\ref{Fig8:TwoObjectDiffPlanes} shows the results diagram for a similar problem in which there are two objects, but this time located in two planes at different axial positions with respect to the source.  Thus, the number of estimated parameters increases to six, with the distances to object A and B being designated $\TwoObjectParam{d}{A}$ and $\TwoObjectParam{d}{B}$, respectively.  While the scenarios may be similar, the implementation of multiple object planes is more complicated than that of a single plane, requiring multiple generalized sources located in different planes.  In this case, the objects are sufficiently separated transversely that we can treat the resulting coherence function as the superposition of the individual coherence functions \cite{Sukhov:17}, each source having the same form as in the one-object example.  Specifically, the resulting coherence function $\TwoObjectSum$ is calculated as
\begin{align}
\TwoObjectSum (y_1,y_2) = G^-(y_1,y_2) - \TwoObjectGdapproxInverted{A}(y_1,y_2) - \TwoObjectGdapproxInverted{B}(y_1,y_2),
\end{align}
where $G^-$ is calculated from \eqref{eqn:GaussSchell} with distance $d\!=\!d_0\!+\!\TwoObjectParam{d}{A}$, and $\TwoObjectGdapproxInverted{A}$ and $\TwoObjectGdapproxInverted{B}$ are ``inverted'' coherences due to objects A and B, respectively.  The inverted coherences are calculated using transmission function
$1 \!-\! T_{j,j}$ in place of $T_{j,j}$ in \eqref{eqn:mainresult1}.
Due to the independence of the two generalized sources located at A and B, \eqref{eqn:gradient} can be applied to each without modification.
As shown in Fig.~\ref{Fig8:TwoObjectDiffPlanes}(b), the number of sample points has been increased to include off-axis measurements, i.e. including points with $y_1 \ne 0$, to aid in estimation of the two distances.
As with the five-parameter example, the maximum parameter estimate error is less than 1.2\%.

\section{Object recovery from experimental data} \label{section:ResultsExperiment}

\begin{figure}[t!]
\centering
\includegraphics[scale=1]{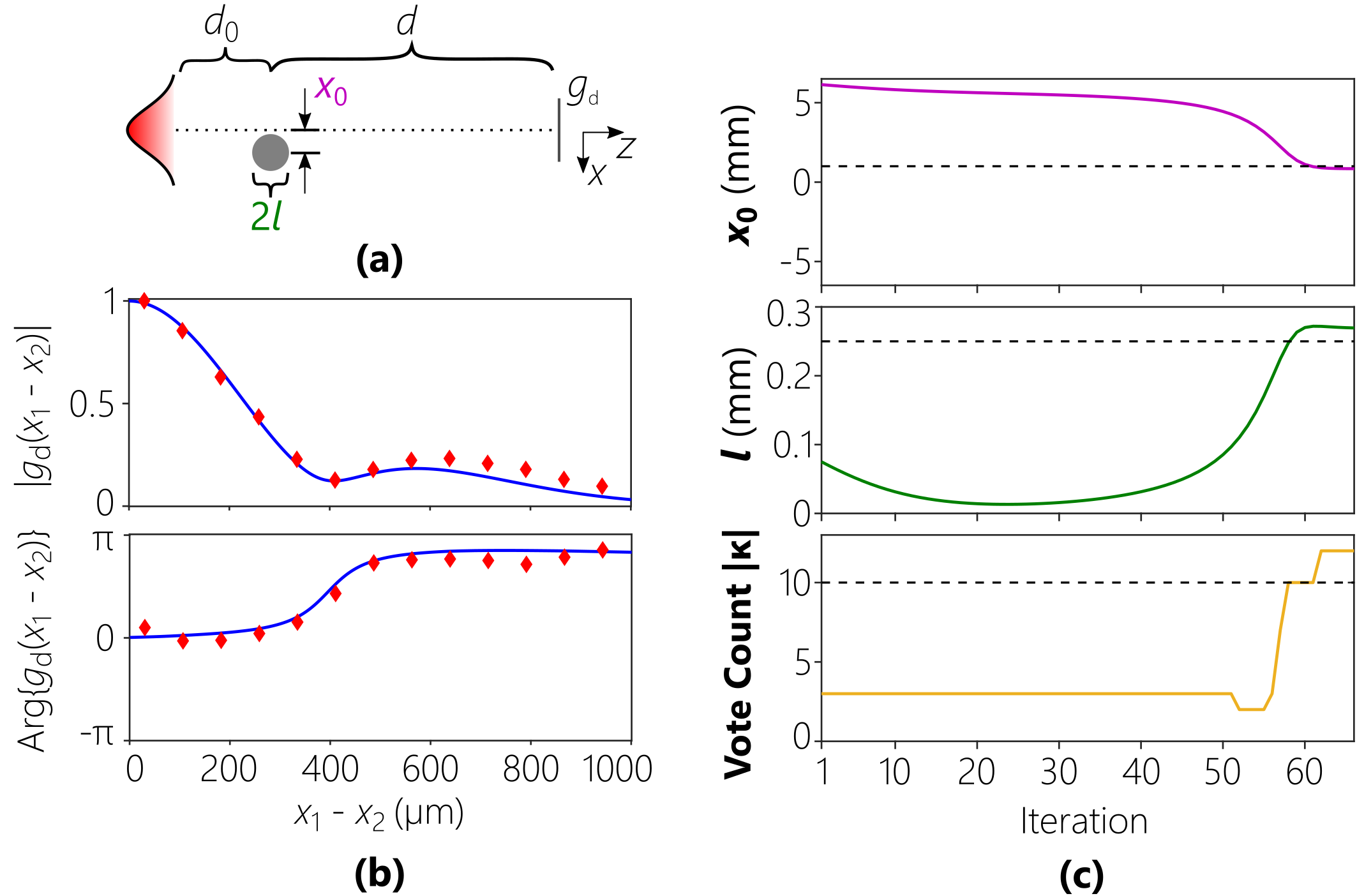}
\caption{
Results of gradient descent algorithm using experimental data.
(a) Diagram of setup.
(b) Modulus and phase of coherence function at measurement plane. Both measured samples and final estimate are shown.
(c) Path of the gradient descent algorithm.  The top two plots show the estimates of the two parameters, with dashed lines indicating the actual value of the parameters.  The bottom plot shows the cardinality of the ``vote'' set $|\ErrorVoteSet|$ at each iteration,
with the threshold $\ThresholdErrorVotes$ indicated by a dashed line.
}
\label{Fig9:OneObjectExperiment}
\end{figure}

In this section, we present results obtained by applying the algorithm to actual experimental measurements from \cite{Kondakci17OE}.  The setup is diagrammed in Fig.~\ref{Fig9:OneObjectExperiment}(a). The source in the experimental setup is a Thorlabs M625L3 LED (with a peak wavelength of $\approx 633$~nm and FWHM-bandwidth of $\approx 18$~nm), with a band-pass filter centered at 632.8 nm and having a bandwidth of $\approx 1.3$-nm FWHM.  The object is a $500 \;\mu$m wire placed at various transverse positions.  The coherence is measured via double slit interferometry by a Digital Micromirror Device (TI DLP6500), a CCD camera (The
ImagingSource, DFK 31BU03), and a set of three lenses (for magnification and to obtain a Fourier transform).

The parameters used for the analytic model are as follows.  The source parameters are $A\!=\!1$, intensity FWHM of 1 mm, coherence FWHM of 75 $\mu$m, and no phase, and the wavelength $\lambda=633$~nm.
The actual object half-width is $l\!=\!0.25$~mm, with varying center $x_0$, and the actual distances are $d_0\!=\!5$~mm, $d\!=\!1.245$~m.

In order to accommodate noise and mismatches in the model, we relax the stopping condition to use a voting mechanism based on the set
\begin{align}
\ErrorVoteSet = \left\{k \,\, \middle| \,\,
|\Gerror(\sampleyI{k},\sampleyII{k})| < \ThresholdError,
\,
1 \le k \le \SampleCount
\right\}.
\end{align}
Specifically, rather than requiring that the residual be small for all samples, here the residual only needs to be small for a subset of the samples.  Additionally, to ensure that individual errors are not excessively large, an additional threshold is placed on
$ \GerrorSquaredSum $.
Accordingly, we replace the condition in \eqref{eqn:GlobalMinCheck} with condition
\begin{align}
\left| \ErrorVoteSet \right| \ge \ThresholdErrorVotes \text{ and }
\GerrorSquaredSum < \ThresholdMeanError
\end{align}
where set cardinality is denoted by $| \! \cdot \! |$.
In this example, the algorithm parameters are set to $\StepSizBreakpoint\!=\!5 \times 10^{-7}$, $\ThresholdGradBreakpoint\!=\!1$,
$\ThresholdError\!=\!0.15$, $\ThresholdErrorVotes\!=\!55$, and $\ThresholdMeanError\!=\!1$.

The resulting estimates of the algorithm generated for several experimental setups are shown in Table \ref{tab:realdata2d}.
The final initialization value is also listed to demonstrate that the algorithm converges given diverse initialization conditions.
To show the low sampling requirements of the proposed algorithm, only 13 of the measured data points are used for estimation.
The detailed gradient descent results for $x_0\!=\!100$~$\mu$m are shown in Fig.~\ref{Fig9:OneObjectExperiment}.
The measured and estimated coherences are shown in Fig.~\ref{Fig9:OneObjectExperiment}(a).
The errors are due to noise in the measurements and inaccurate assumptions in modeling the source as a Gauss-Schell source.
The gradient descent dynamics are shown in Fig.~\ref{Fig9:OneObjectExperiment}(b).  Rather than showing all initializations, as was done in Section \ref{section:Results}.\ref{section:ResultsOneObject}, only the final initialization is shown (i.e., the successful initialization which converges to the global minimum).  As seen in the bottom plot of Fig.~\ref{Fig9:OneObjectExperiment}(b), only when the parameters approach the actual values does the residue become small, and we have $\left| \ErrorVoteSet \right| \ge \ThresholdErrorVotes$.

\begin{table}[t!]
  \centering
  \caption{\bf Experimental results. For each parameter, the actual value, initial starting point, and final estimate are shown. }
  \begin{tabular}{ccc ccc}
  \hline
    \multicolumn{3}{c}{Center $x_0$ ($\mu$m)}
      & \multicolumn{3}{c}{Width $2l$ ($\mu$m)} \\
    \cmidrule(lr){1-3} \cmidrule(lr){4-6}
    Actual & Initial & Est.
      & Actual & Initial & Est. \\
      \cmidrule(lr){1-3} \cmidrule(lr){4-6}
\hline
-100 & -409.2 & -114.7 & 500 & 137.6 & 479.7 \\                        
-50 & -409.2 & -36.1 & 500 & 137.6 & 555.9 \\                          
0 & 409.2 & 5.6 & 500 & 299.7 & 610.3 \\                               
50 & -306.9 & 57.3 & 500 & 58.4 & 583.7 \\                             
100 & 613.9 & 84.5 & 500 & 75.0 & 540.7 \\    
    \hline
  \end{tabular}
  \label{tab:realdata2d}
\end{table}

\section{Discussion}\label{section:Discussion}
We considered the inverse problem of recovering the location and shape of a modulated, partially coherent Gauss-Schell source propagating in the Fresnel regime, by using coherence measurements \cite{Beckus:17}. 

We introduced a global minimum-residual inversion method that relies on the closed-form coherence formula derived recently by the authors \cite{Beckus:17}. For presentation purposes, the minimization problem is solved by a simple gradient descent algorithm, which builds on prior information where available.
More sophisticated algorithms, e.g., the Levenberg–Marquardt \cite{Seber:89}, could be used in the minimization problem to improve the convergence rate.

In applications with simulated data, we demonstrated that the method determines the size and location of intercepting single and double objects, even when they are located in separate transverse planes.

The reconstruction method is robust and works well also with experimental data, as presented in Section 5 above.

Whereas we applied the method only to determine the breakpoints and distance (corresponding to multiple obscurants and apertures), it can also be used to estimate a piecewise constant complex-valued transmission function, as well as the statistical parameters of the source.

\begin{figure*}[t!]
\centering
\includegraphics[scale=1]{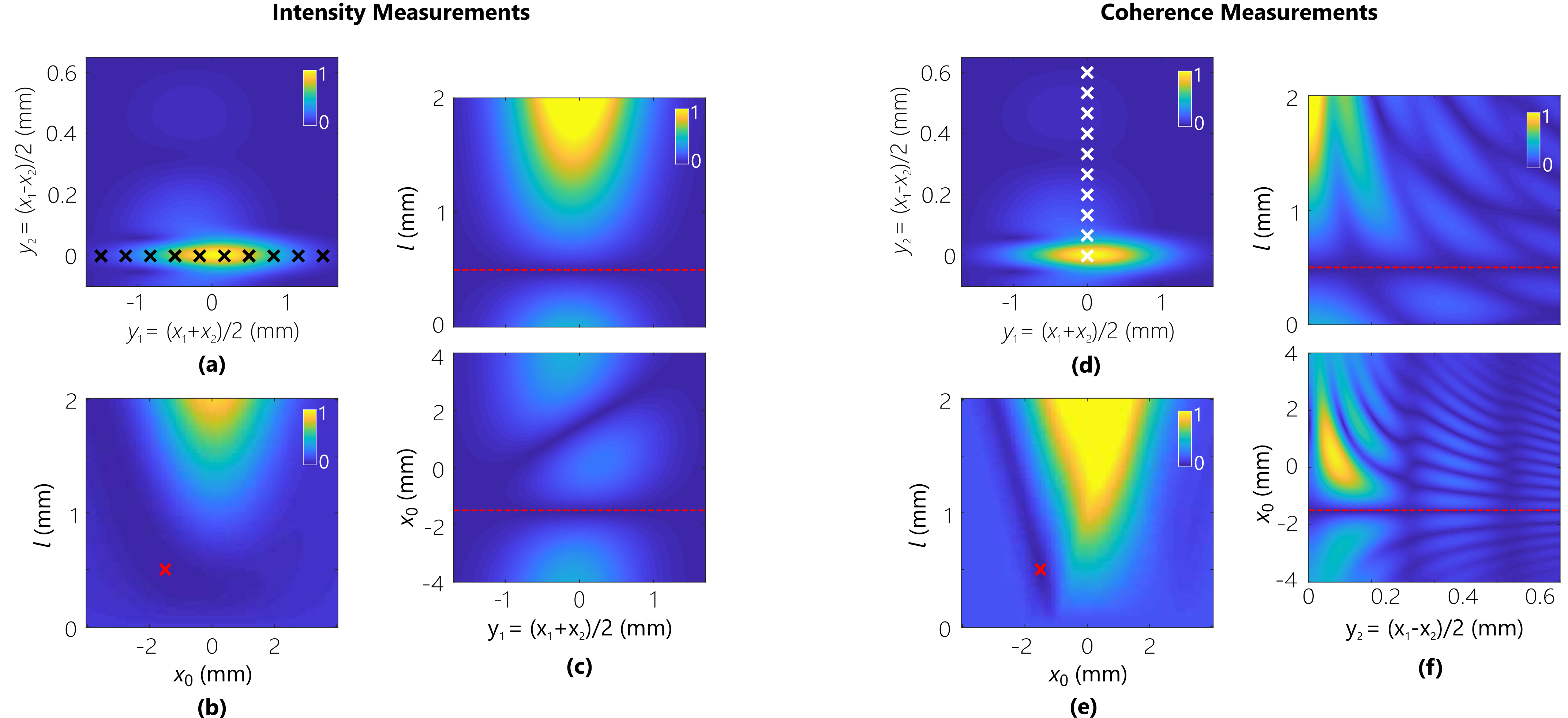}
\caption{
Comparison of intensity and coherence measurements.  The modulus of the simulated coherence function is shown in (a) and (d) with intensity sample points indicated by black ``x'' marks and coherence sample points indicated indicated by white ``x'' marks.  The corresponding residual maps $\GerrorSquaredSum(x_0,l)$ for the two scenarios are shown in (b) and (e).  For comparison purposes, the functions are normalized against $\frac{1}{\SampleCount} \sum_{k=1}^M | G_d(y_1^k,y_2^k) |^2$, and plotted on the same scale.
As can be seen in this example, the residual map for intensity measurements exhibits a larger area of minima than that of the coherence measurements.  This may lead to more ambiguity in the reconstruction, although results will vary depending on physical factors such as the signal-to-noise ratio of the measurements.
(c) and (e) show the residual $\Gerror$ plot as a function of the sample point (along the horizontal) and parameter (vertical).  Each plot shows variation with regard to one parameter while the other is fixed at the correct value, and all plots use the same scale.  The actual parameter values are indicated in red.  The parameters are the same as used in Fig.~\ref{Fig3:OneObjectDescent}.
}
\label{Fig10:IntVsCoh}
\end{figure*}

The coherence measurements bring in an additional dimension to the data, which allows for devising a global inversion method. More precisely, the local method of steepest descent is applied to a family of residuals, all of which have a common unique minimizer. 
This idea is stressed in the example in Fig.~\ref{Fig10:IntVsCoh}, where the residuals are calculated for a family of functions (parameterized along the vertical axes), by using the sample points along the horizontal axis.  
The global minimum is the unique point at which all these functions are zero.

\appendix

\section {Derivation of Gradients} \label{section:DeriveGradients}
From \eqref{eqn:ConjugatedHilbertTransform} and \eqref{eqn:mainresult1}, assuming $T_{1,1}=T_{N+1,N+1}=0$,
\begin{align} \label{eqn:derivation1}
\Gdapprox(y_1,y_2)
&= \widetilde{G}_d(y_1,y_2) \frac{i}{2 \Gaussian{\eta \widetilde{\sigma}}{y_2}} \sum_{j=1}^N (T_{j,j}-T_{j+1,j+1})
\nonumber \\
&\quad \times \exp \left\{ -i y_2 b_j(y_1) \right\}
\nonumber \\
&\quad \times p.v. \frac{1}{\pi} \int \frac{\exp \left\{ i s b_j(y_1) \right\} \Gaussian{\tsigma / \eta}{s}}{y_2-s} ds
\end{align}
Then,
\begin{align}
\Gdapprox(y_1,y_2)
&= \frac{i \eta^3 \ell^2}{\sqrt{2 \pi} \tsigma} \GradConst(y_1,y_2) \sum_{j=1}^N (T_{j,j}-T_{j+1,j+1})
\nonumber \\
&\quad \times \exp \left\{ -i y_2 b_j(y_1) \right\}
\nonumber \\
&\quad \times p.v. \frac{1}{\pi} \int \frac{\exp \left\{ i s b_j(y_1) \right\} \Gaussian{\tsigma / \eta}{s}}{y_2-s} ds
\nonumber \\
&= \frac{i \eta^3 \ell^2}{\sqrt{2 \pi} \tsigma} \GradConst(y_1,y_2) \sum_{j=1}^N (T_{j,j}-T_{j+1,j+1})
\nonumber \\
&\times p.v. \! \int \frac{\exp \left\{ i (s-y_2) b_j(y_1) \! + \! i y_{1}y_{2}/\tR^{2} \right\} \Gaussian{\tsigma / \eta}{s}}{y_2-s} ds,
\end{align}
where
\begin{align}
\GradConst(y_1,y_2)
=& \frac{\tA \tsigma }{\sqrt{ 2 \pi } \eta^3 \ell^2} \Gaussian{\tw}{y_{1}} \Gaussian{\ell^2 \eta / \sigma}{y_2}.
\end{align}
The partial derivative of the real component of $\Gdapprox$ with respect to breakpoint $a_j$ is
\begin{align} \label{eqn:derivation2}
\frac{\partial \realpart{\Gdapprox}}{\partial a_j}
=&  (T_{j,j}-T_{j+1,j+1}) \frac{1}{\sqrt{2 \pi} \tsigma} \GradConst(y_1,y_2)
\nonumber \\
&\times \cos \left( y_2 b_j(y_1) - y_{1}y_{2}/\tR^{2} \right)
\nonumber \\
&\times \int \cos \left( s b_j(y_1) \right) \Gaussian{\tsigma / \eta}{s} ds.
\end{align}
Using the definition
\begin{align}
\GradGdapproxExp(y_1,y_2) =& (T_{j,j}-T_{j+1,j+1}) \GradConst(y_1,y_2)
\nonumber \\
&\times \exp \left( i y_2 b_j(y_1) - i y_{1}y_{2}/\tR^{2} \right) \GaussianNoParm{\eta / \tsigma} \left( b_j(y_1) \right),
\end{align}
we can express \eqref{eqn:derivation2} as
\begin{align}
\frac{\partial \realpart{\Gdapprox}}{\partial a_j}
=&  (T_{j,j}-T_{j+1,j+1}) \GradConst(y_1,y_2)
\nonumber \\
& \times \cos \left( y_2 b_j(y_1) - y_{1}y_{2}/\tR^{2} \right) \GaussianNoParm{\eta / \tsigma} \left( b_j(y_1) \right)
\nonumber \\
=& \realpart{\GradGdapproxExp(y_1,y_2)}
\end{align}
Similarly, for the imaginary part,
\begin{align}
\frac{\partial \imagpart{\Gdapprox}}{\partial a_j} = -\imagpart{\GradGdapproxExp(y_1,y_2)}
\end{align}

We now introduce an arbitrary normalization function $\GradNorm(y_1,y_2)$, and consider the residual function
\begin{align}
\GerrorGeneral(y_1,y_2;\avec,d) = \frac{\Gdapprox(y_1,y_2;\avec,d)}{\GradNormApprox(y_1,y_2)} - \frac{G_d(y_1,y_2)}{\GradNorm(y_1,y_2)}
\end{align}
For the remainder of this section, to facilitate readability, the function parameters $(y_1,y_2)$ will be omitted.  The partial derivative of the squared modulus of $\GerrorGeneral$ is
\begin{align} \label{eqn:derivabsfsquared_start}
\frac{\partial}{\partial a_j} |\GerrorGeneral|^2
=& 2 \realpart{\GerrorGeneral} \frac{\partial \realpart{\GerrorGeneral}}{\partial a_j} 
+ 2 \imagpart{\GerrorGeneral} \frac{\partial \imagpart{\GerrorGeneral}}{\partial a_j}  \nonumber \\
=& 2 \realpart{\GerrorGeneral} \frac{\frac{\partial \realpart{\Gdapprox}}{\partial a_j} \GradNormApprox - \frac{\partial \GradNormApprox}{\partial a_j} \realpart{\Gdapprox}}{\GradNormApprox^2}
\nonumber \\
& + 2 \imagpart{\GerrorGeneral} \frac{\frac{\partial \imagpart{\Gdapprox}}{\partial a_j} \GradNormApprox - \frac{\partial \GradNormApprox}{\partial a_j} \imagpart{\Gdapprox}}{\GradNormApprox^2}
\nonumber \\
=& \frac{2}{\GradNormApprox}
\left\{ \realpart{\GerrorGeneral} \left( \realpart{\GradGdapproxExp}
- \frac{\partial \GradNormApprox}{\partial a_j} \realpart{\frac{\Gdapprox}{\GradNormApprox}} \right) \right.
\nonumber \\
& - \left. \imagpart{\GerrorGeneral} \left( \imagpart{\GradGdapproxExp} - \frac{\partial \GradNormApprox}{\partial a_j} \imagpart{\frac{\Gdapprox^*}{\GradNormApprox}} \right) \right\}.
\end{align}
For unnormalized coherence, substituting $\GradNormApprox=1$ yields
\begin{align} \label{eqn:derivabsfsquared_unnorm}
\frac{\partial}{\partial a_j} |\Gerror|^2
=& 2 \left\{ \realpart{\Gerror} \realpart{\GradGdapproxExp}
- \imagpart{\Gerror} \imagpart{\GradGdapproxExp} \right\}
\nonumber \\
=& \sqrt{\frac{2}{\pi}} \frac{\tA \tsigma}{\eta^3 \ell^2}
\left( T_{j,j} - T_{j+1,j+1} \right)
\nonumber \\
& \times \left\{ \realpart{\Gerror} \, \realpart{\GradMain{j}}
- \imagpart{\Gerror} \, \imagpart{\GradMain{j}} \right\}
\end{align}
with $\GradMain{j}$ as defined in \eqref{eqn:gradmain}.  The summation \eqref{eqn:gradient} follows immediately.

For the degree of coherence, substituting the normalization $\GradNormApprox=\sqrt{\IdapproxI \IdapproxII}$ gives the residual $\gerror$ defined in \eqref{eqn:pointerror_normalized}.  Note that the normalized coherence is defined to be zero if either of the intensities is zero.  The partial derivative of the normalization term is
\begin{align} \label{eqn:GradientNormalization_gd}
&\frac{\partial \sqrt{\IdapproxI \IdapproxII}}{\partial a_j}
\nonumber \\
&= \frac{1}{2} \left( \IdapproxI \IdapproxII \right)^{-1/2} \left[ \IdapproxI \frac{\partial \Gdapprox(y_1-y_2,0)}{\partial a_j}
+ \IdapproxII \frac{\partial \Gdapprox(y_1+y_2,0)}{\partial a_j} \right]
\nonumber \\
&= \frac{T_{j,j}-T_{j+1,j+1}}{2 \sqrt{ \IdapproxI \IdapproxII } } \left\{ \IdapproxI \GradConst(y_1-y_2,0)  \GaussianNoParm{\eta / \tsigma} \left( b_j(y_1-y_2) \right)
\right.
\nonumber \\
&\quad\left. + \IdapproxII \GradConst(y_1+y_2,0) \GaussianNoParm{\eta / \tsigma} \left( b_j(y_1+y_2) \right) \right\}
\end{align}
Substituting \eqref{eqn:GradientNormalization_gd} into \eqref{eqn:derivabsfsquared_start} and continuing,
\begin{align} \label{eqn:derivabsfsquared_norm}
\frac{\partial}{\partial a_j} |\gerror|^2
=& \frac{2}{\sqrt{\IdapproxI \IdapproxII}}
\left\{ \realpart{\gerror} \left( \realpart{\GradGdapproxExp} - \frac{\partial \sqrt{\IdapproxI \IdapproxII}}{\partial a_j} \realpart{\gdapprox} \right)
\right.
\nonumber \\
& \left. - \imagpart{\gerror} \left(\imagpart{\GradGdapproxExp} - \frac{\partial \sqrt{\IdapproxI \IdapproxII}}{\partial a_j} \imagpart{\gdapprox^*} \right) \right\}
\nonumber \\
=& \sqrt{\frac{2}{\pi}} \frac{\tA \tsigma}{\eta^3 \ell^2}
\left( T_{j,j} - T_{j+1,j+1} \right)
\nonumber \\
& \times \left\{ \realpart{\gerror} \, \realpart{\GradMainNorm{j}}
- \imagpart{\gerror} \, \imagpart{\GradMainNorm{j}} \right\}
\end{align}
with $\GradMainNorm{j}$ as defined in $\eqref{eqn:gradmain_normalized}$.

\section*{Funding.} DARPA under contract HR0011-16-C-0029

\bibliography{main}

\end{document}